\documentclass[aps,prb,preprint,superscriptaddress,11pt]{revtex4-1} 

\usepackage{amsmath,amssymb}
\usepackage{graphicx}
\usepackage{epsfig}
\usepackage{color}

\begin{document}

\title{High-sensitivity high-speed compressive spectrometer for Raman imaging}

\author{Benneth Sturm}
\address{D\'{e}partement de Physique, Ecole Normale Sup\'{e}rieure / PSL Research University, CNRS, 24 rue Lhomond, 75005 Paris, France}
\author{Fernando Soldevila}
\address{GROC-UJI, Institute of New Imaging Technologies (INIT), Universitat Jaume I, E12071 Castell\`o, Spain}
\author{Enrique Tajahuerce}
\address{GROC-UJI, Institute of New Imaging Technologies (INIT), Universitat Jaume I, E12071 Castell\`o, Spain}
\author{Sylvain Gigan}
\address{Laboratoire Kastler Brossel, CNRS UMR 8552, Ecole Normale Sup\'{e}rieure, PSL Research University Sorbonne Universit\'{e}s \& Universit\'{e} Pierre et Marie Curie Paris 06, F-75005, Paris, France}
\author{Herve Rigneault}
\address{Aix Marseille Univ, CNRS, Centrale Marseille, Institut Fresnel, F-13013 Marseille, France}
\author{Hilton B. de Aguiar}
\email{Corresponding author: h.aguiar@phys.ens.fr}
\address{D\'{e}partement de Physique, Ecole Normale Sup\'{e}rieure / PSL Research University, CNRS, 24 rue Lhomond, 75005 Paris, France}

\begin{abstract}%
Compressive Raman is a recent framework that allows for large data compression of microspectroscopy during its measurement. Because of its inherent multiplexing architecture, it has shown imaging speeds considerably higher than conventional Raman microspectroscopy. Nevertheless, the low signal-to-noise (SNR) of Raman scattering still poses challenges for high-sensitivity bio-imaging exploiting compressive Raman: (\textit{i}) the idle solvent acts as a background noise upon imaging small biological organelles, (\textit{ii}) current compressive spectrometers are lossy precluding high-sensitivity imaging. We present inexpensive high-throughput spectrometer layouts for high-sensitivity compressive hyperspectroscopy. We exploit various modalities of compressive Raman allowing for up to 80X reduction of data storage and 2X microspectroscopy speed up at a 230-nm spatial resolution. Such achievements allowed us to chemically image sub-diffraction-limited biological specimens (lipid bilayers) in few seconds.
\end{abstract}

\maketitle

\section{Introduction}
Raman microspectroscopy allows probing a chemically heterogeneous system with high spatial resolution and superb molecular selectivity. In particular, Raman-based imaging holds promising applications for biological specimens studies~\cite{Kong2015}, such as for cancer~\cite{Harmsen2015,Jermyn2015} and bacteria detection~\cite{Lorenz2017,Bodelon2016}, molecular biology imaging~\citep{Okada2012,Klein2012,Ando2015,Donaldson2018}, to cite few examples. Nevertheless, hyperspectroscopy (\textit{i.e.} spatially-resolved spectroscopy), in general, generates large data sizes which will eventually become a hurdle when bringing Raman spectroscopic imaging as a valuable tool for clinical tests, for example, for issues related to data storage and access. As Raman microspectroscopy acquisition speeds continuously advance~\cite{Shipp2017}, it will become increasingly important to achieve data compression, in particular for dynamic and/or large-scale imaging at high-resolution.

Single-pixel-based compressive sensing technology has attracted a large interest recently~\cite{Duarte2008}. The idea behind compressive sensing is to obtain the same information as in multi-pixel technologies approaches, however through a designed optical experiment that will compress the data at the acquisition stage, by undersampling. In a post-processing step, efficient computational methods retrieve the desired information accurately. Interestingly, single-pixel approaches are particularly powerful where multi-pixel technology is still lacking or challenging, such as in terahertz and pump-probe spectroscopy methods~\cite{Berto2017,Chan2008}. Single-pixel compressive sensing is a very appealing approach for Raman imaging as it is inexpensive and may have very fast computational reconstruction times, allowing for real-time applications. Compressive microspectroscopy is based on spectral, or spatio-spectral multiplexing~\cite{August2013}, which is achieved by fast programmable spatial light modulators, such as digital micromirror devices (DMD), operating at 10's kHz.

While compressive microspectroscopy~\cite{Davis2011,Galvis-Carreno2014,Thompson2017,Wilcox2012,Wilcox2013,Wei2016,Refregier2018, Scotte2018,Corcoran2018,August2013,Pavillon2016,Wadduwage2017,Zhang2018} is a promising technology, it still lacks sensitivity for bio-imaging when compared to sensitive CCD cameras. High-sensitivity imaging is of particular relevance in the sub-micron spatial scale, as it is the size range of many organelle in a cell~\cite{Klein2012,Ando2015,Rieger2017,Donaldson2018}, including lipid droplets, cell membranes, and membrane-less organelles. The challenges for high-sensitivity compressive microspectroscopy are two-fold. First, apart from Raman scattering being a weak effect, the Raman signal ($I_R$) scales unfavourably with spatial resolution: $I_R\propto\omega_0^2$ for objects that fill the focus ($\omega_0$ being the focus size of the excitation laser). Therefore, upon low-resolution imaging, the overall signal increases but the \textit{sensitivity} decreases for detecting objects that have a size smaller than the focus, as the surrounding solvent adds as a strong background. Second, and related to compressive spectrometer layouts, there is a strong size mismatch (100X) between sensitive detector sizes and the size of the spectrum spread in the Fourier plane of a spectrometer (i.e. on the DMD). This mismatch is the main cause of sensitivity loss in compressive spectrometers. 

Here, we exploit novel high-throughput and inexpensive spectrometer designs for high-resolution and high-sensitivity compressive Raman imaging of biological specimens. This is achieved by dispersion-controlled layouts to focus a large spectral bandwidth on a small single-pixel detector. After discussion of the methods we propose, we showcase two compressive Raman modalities exploiting different levels of a priori knowledge from the system being imaged. We then demonstrate chemical imaging of sub-wavelength Raman sources using the compressive spectrometer.

\section{Methods}
\subsection*{Confocal microscope}
The compressive Raman microscope is based on a standard layout of an epi-detection Raman microscope (Fig.~\ref{fig1}). The excitation laser ($\lambda=532$~nm, 150~mW, Oxxius LCX-532) is cleaned up by a bandpass filter (FLH05532-4, Thorlabs), and expanded to about 7.2~mm, before entering an inverted microscope (Nikon Eclipse Ti-U). An additional half-waveplate and quarter-waveplate (not shown) pre-compensate and generate circularly polarized light~\cite{Belanger2009} after the beam being reflect off from the dichroic filter (F38-532\_ T1, AHF) into a high-numerical aperture (NA), oil-immersion objective (Nikon 60X, 1.4~NA). Specimens are mounted on a translation-stage (Physik Instrumente P-545.3R7) for raster scanning. The inelastically backscattered Raman light (epi configuration) is focussed by the microscope tube lens on a multimode fiber (M50L02S-A, Thorlabs, 50~$\mu$m core-size), with a notch filter blocking residual pump light (NF533-17, Thorlabs), thus guiding the signal to the home-built compressive spectrometer.

\subsection*{Single-pixel compressive spectrometer}
The home-built spectrometer is based on a traditional Czerny-Turner design. The signal coming out of the fiber is collimated by an achromatic lens L1 (f=19mm, AC127-019-A-ML, Thorlabs), which then impinges on a high-efficiency transmission grating (FSTG-VIS1379-911, Ibsen Photonics, 1379~l/mm) providing about 85\,\% diffraction efficiency around $\lambda$=600~nm. After the grating, the beam is spectrally dispersed and different colors are focused with an off-axis parabolic mirror PM (f=101.6~mm, 90$^\circ$ off-axis angle, MPD149-P01, Thorlabs). The DMD module (V-7001, ViALUX, 0.7" XGA resolution, with V4395 controller board) selects the various wavelength to be detected. Typically, we bin about 8 pixels, thus having a macropixel size that is slightly below the spectral resolution of the spectrometer. 

The signal from the DMD is steered into an avalanche photodiode operating in photon counting mode (SPCM-AQRH-44, Excelitas Technologies). This module combines high quantum efficiency from Si avalanche photodiodes (70~\%), large dynamic range (low dark and high maximum count rate), with an active area diameter of 180~$\mu$m. The small area, while keeping high quantum efficiency, is of particular importance in the design because focusing the complete spectrum to the size of the detector-chip efficiently is not straightforward. We note other single-photon counters exist with mm's active area size, however, with lower quantum efficiency.

\subsection*{Sample preparation}
Polystyrene particles (1~$\mu$m and 0.088~$\mu$m-diameter, Polysciences Inc.) were drop cast on a coverslip, then filled with water. Multilamellar vesicles (MLV) were prepared by drop-casting a film of lipid solution (10~mg/ml dipalmitoylphosphatidylcholine in chloroform) on a vial, then adding water and heat up at $38^\circ$ for about 30~min~\cite{Gasecka2017}. After spontaneous formation of the vesicles, an aliquot of the dispersion is then sandwiched between two coverslips.

\begin{figure}[htbp]
	\centering{
	\includegraphics[width=0.75\columnwidth]{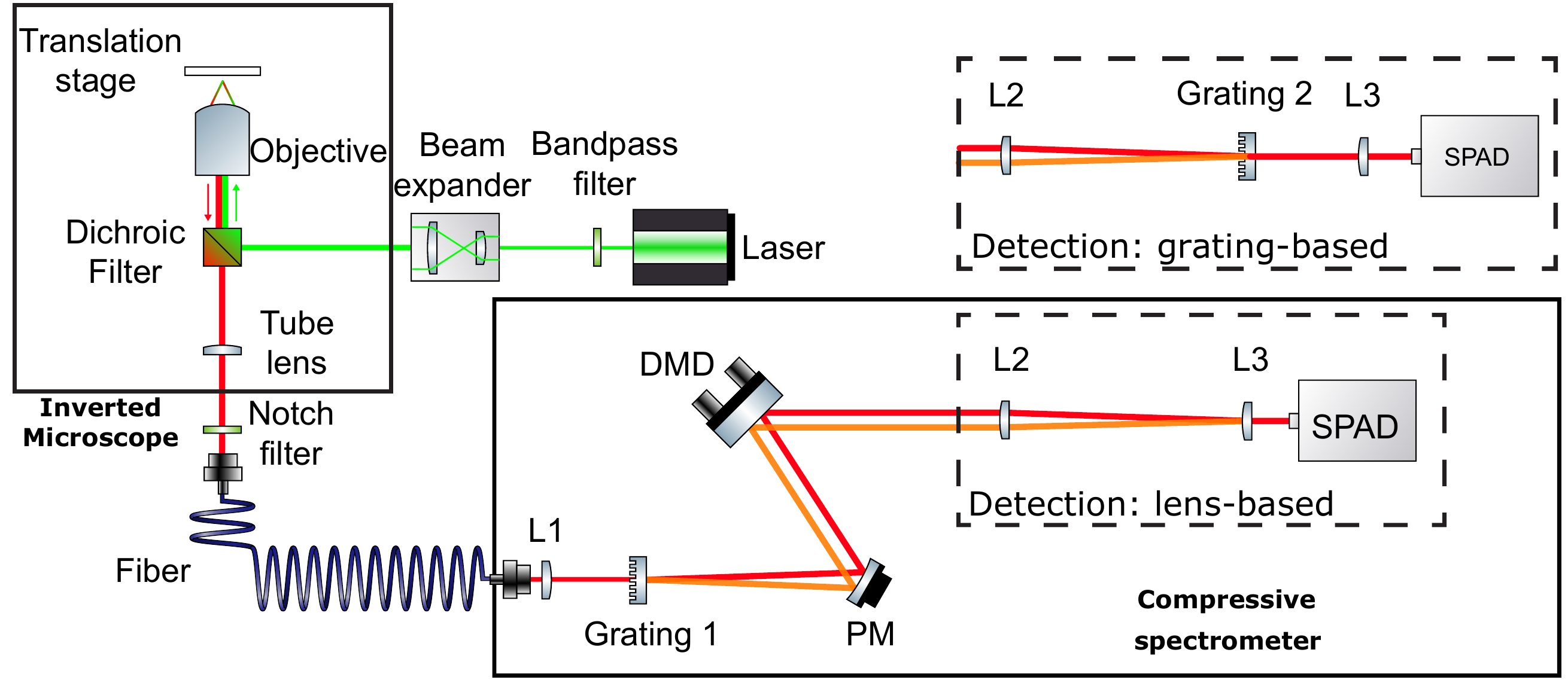}}
	\caption{
		Simplified layout of the compressive Raman microscope. A CW laser source excited Raman scattering process, which is detected in epi configuration. The sample is raster-scanned with a high-resolution piezo-scanner. The Raman inelastically scattered light is steered into a home-built spectrometer designed for compressive single-pixel experiments. The dashed boxes depicts two strategies (grating or lens telescope) for dispersion management, introduced by the first grating, to focus large bandwidths in the high sensitivity single-pixel detector. DMD=digital micromirror device, PM=off-axis parabolic mirror, L=lens.		
	}
	\label{fig1}
\end{figure}

\section*{Results}
\subsection*{Dispersion management for single-pixel compressed spectroscopy}
The main challenge of a single-pixel compressive spectrometer is as follows. In a conventional spectrometer, a CCD camera is used to detect the dispersed input light, which size is typically few mm's. In contrast, the compressive spectrometer consists of a DMD, in place of the CCD, which steers wavelength bins onto a small single-pixel detector (100's~$\mu$m) to achieve high-sensitivity. Therefore, one has to considerably demagnify the beam reflected off by the DMD. Hence, we studied two strategies that could keep the interesting aspects (costs and throughput) of single-pixel detection. 


We investigated two approaches that are presented in dashed box insets of Fig.~\ref{fig1}. The first approach is based solely on lenses (lens-based), motivated by the fact that transmission of lenses are highest compared to other optical elements. To aid the discussion, we recall the \textit{Smith-Helmholtz invariant} ($y\sigma = y'\sigma'$) relating planes~$y$ (grating 1), ray angles~$\sigma$ (arising from grating dispersion before PM), to $y'$ (L3) and $\sigma'$ (arising from focusing with L2) of an optical system. As the grating plane $y$ is imaged onto plane $y'$ by a PM and L2, we have:
\begin{equation}
\frac{\sigma}{\sigma'} = \frac{y'}{y} = \frac{f_2}{f_1} = m.\label{eq1}
\end{equation}
Eq.~\ref{eq1} shows that dispersions and magnifications can be interchanged. For instance, $\sigma'$ can be reduced by simply increasing the magnification $m$. However, the maximum magnification is limited by the clear aperture of L3. The second approach is based on a second grating to cancel the dispersion of the first one (Fig.~\ref{fig1}, grating-based), as typically done in pulse shapers of ultrafast spectroscopy, if they have the same groove density. Note that if the second grating is of different dispersion, one can adjust $f_2$ to match the dispersion given by $\sigma'$. 

We have compared the two methods. For the lens-based approach, we used an effective 4X demagnification (in the figure it is presented as a single lens L2). For the grating-based method, we used a second grating with the same dispersion as the first grating. We have inspected the spectrum of the lens-based approach (17~nm-bandwidth) and grating-based approach (77~nm-bandwidth). While both methods coupled large bandwidths (over 100's~cm$^{-1}$), the lens-based had too much losses originating from clipping in the edge of lenses. Therefore, we used the grating-based design. Nevertheless, we believe the lens-based throughput could overcome the grating-based if a low-dispersion grating is used.

\subsection*{Compressive Raman without a priori knowledge}
We exploited the spectrometer for two different types of compressive Raman schemes. Basically, there are two ways of performing compressive Raman: with (supervised)~\cite{Wilcox2012,Wilcox2013,Scotte2018,Wei2016,Corcoran2018} or without (unsupervised)~\cite{Galvis-Carreno2014,Thompson2017} a priori knowledge about the principal components (or eigenvectors) of the hyperspectrum. In this section, we present compressive Raman results without a priori knowledge. 

\begin{figure}[t!]
	\centering{
	\includegraphics[width=0.75\columnwidth]{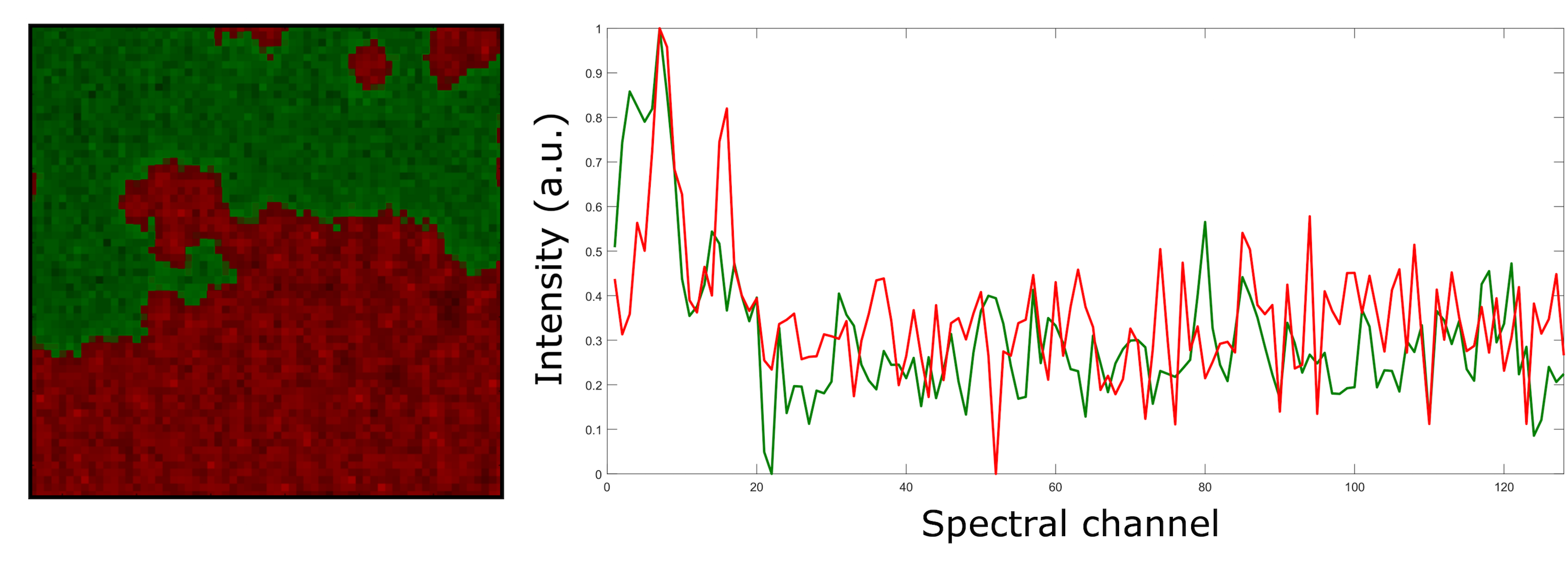}}
	\caption{
	Compressive Raman imaging without a priori knowledge. Left: Raman pseudocolored image of polystyrene beads obtained by means of computational reconstruction. Right: Spectral content of both species present in the sample, water (green), and polystyrene (red). Compression: 2X.
	}
	\label{fig3}
\end{figure}

The measurement process of the hyperspectrum is done by a combined use of the piezo-scanner and the DMD, described elsewhere~\cite{Soldevila2018}. Briefely, the sample is illuminated in a raster scan manner over the image plane. For each spatial position, the Raman signal generated by the sample is measured using the compressive spectrometer. Instead of performing a wavelength scan over the spectral domain, the spectral content is measured using multiplexed detection with Hadamard functions generated with the DMD. One key advantage of using Hadamard basis is for increasing the SNR~\cite{Decker1971,Deverse2000,August2013} because half the pixels are always on. This sampling scheme can be expressed in matrix form as $\mathbf{y}=\mathbf{S}\mathbf{x}$, where the vector $\mathbf{y}$ contains each one of the integrated measurements, $\mathbf{x}$ is the hyperspectral object (in vector form), and the whole sensing process is described by the linear operator $\mathbf{S}$, which is the Kronecker product of the two sampling bases used here: $\mathbf{S}=\mathbf{I}\otimes\mathbf{H}$ ($\mathbf{I}$ being the canonical spatial-scanning basis, and $\mathbf{H}$ the Hadamard spectral-scanning basis). In order to reduce the number of measurements, we select a random subset of rows of $\mathbf{S}$ to measure the hyperspectral dataset. Then, we use compressive sensing algorithms to solve the following optimization problem:
\begin{equation}
\arg \min \mathbf{\hat{x}} = \frac{1}{2}\Vert\mathbf{y}-\mathbf{S}\mathbf{\hat{x}}\Vert^2_2+\tau\Vert\mathbf{\hat{x}}\Vert_{TV}
\end{equation}
based on the Total Variation as the regularizing function, which has been previously reported to provide good results in multispectral imaging scenarios~\cite{Bioucas-Dias2007,Rajwade2013}. The reconstructions can be seen in Fig.~\ref{fig3} of a relatively small region of the sample (64x64 spatial positions and 128 spectral channels). From the spectral information, we can distinguish between the water content of the sample (coloured green), and the polystyrene beads (red). 

\subsection*{Compressive Raman with a priori knowledge}
\begin{figure}[t!]
	\centering{
	\includegraphics[width=0.75\columnwidth]{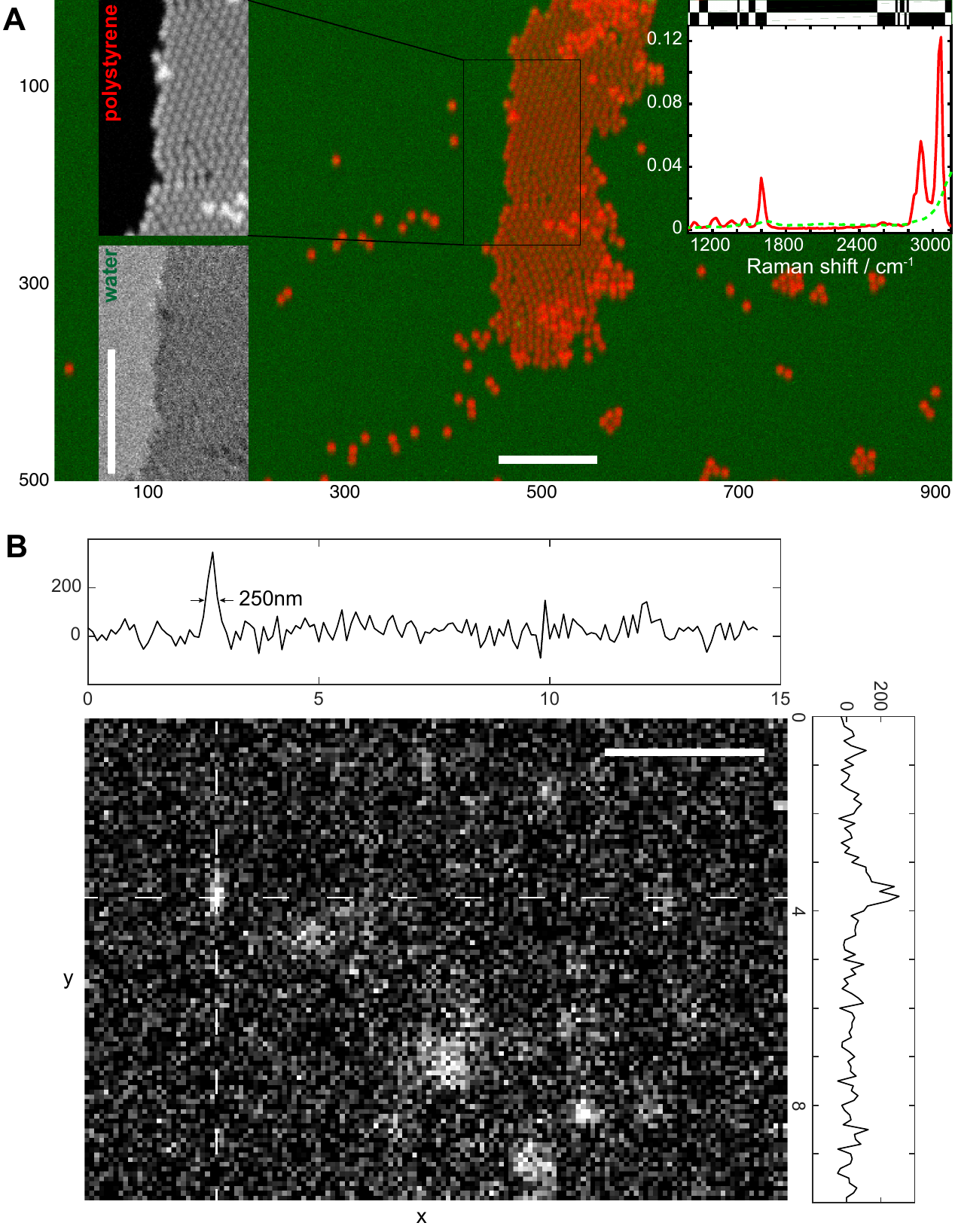}}
	\caption{
		Compressive Raman imaging with a priori detection.
		(A) Large scale imaging with 80X compression (axis are in pixel count). Raman pseudo-color images of 1~$\mu$m-polystyrene beads (red, upper left inset) and water (green, lower left inset). Spectra are presented in the inset with the optimized filters shown on top. Scale bar: 20~$\mu$m. Effective pixel dwell time: 640~$\mu$s.
		(B) High-sensitivity imaging of 88~nm-polystyrene beads. Only the polystyrene channel is presented. Scale bar: 1.5~$\mu$m. Effective pixel dwell time: 3~ms
	}
	\label{fig4}
\end{figure}

In supervised compressive Raman, one estimates the proportions of a linear combination of $M$ eigenspectra ($\mathbf{P}$), from $N$ measurements ($\mathbf{x}$). This is achieved by projecting the input spectrum onto $N$ optimized spectral filters ($\mathbf{F}$) with the DMD. Each spatial pixel can then be written as $\mathbf{x}=\mathbf{F~P^\top}~\mathbf{c}$, where proportions $\mathbf{c}$ are estimated from simple inversion of $\mathbf{F~P^\top}$. The key step for proportion estimation is to develop filters that are robust against noise~\cite{Wilcox2012,Refregier2018}. We use as a metric the variance of the Cramer-Rao bound~(CRB) to determine the optimized spectral filters. A thorough description of the method can be found elsewhere~\cite{Refregier2018,Scotte2018}.

Fig.~\ref{fig4} presents results of large-scale and high-sensitivity chemical imaging. Fig.~\ref{fig4}A shows an example of the compressive detection approach where a large-scale high-resolution image (0.5~Mpixel) is presented with chemical selectivity. After generation of the optimized filters (Fig.~\ref{fig4}A, optimized filters are presented above spectra in the top-right inset), we obtain $\mathbf{x}$ from which we estimate $\mathbf{c}$ in a pixel-wise manner. Chemically selective images are then presented in pseudo-color images (Fig.~\ref{fig4}A) or analysed separately in their respective channels (Fig.~\ref{fig4}A insets shows water and polystyrene proportions). Due to high-resolution imaging, the pseudo-color images are highly contrasted (lack of yellow color).

To further quantify the resolution and demonstrate high-sensitivity detection, we present images of 88-nm polystyrene beads, sizes below the diffraction limit of the system, in Fig.~\ref{fig4}B. Despite the low SNR, one observes regions of aggregated beads with high intensities, and isolated ones with much lower intensities. To estimate the resolution of the system, we use a simple deconvolution using the measured line profiles of the low-intensity beads ($\approx$250~nm). We obtain a resolution of $\approx$230~nm, a value close to the diffraction limit of the system.

\subsection*{Sensitivity to model membranes}
\begin{figure}[t!]
	\centering{
	\includegraphics[width=0.75\columnwidth]{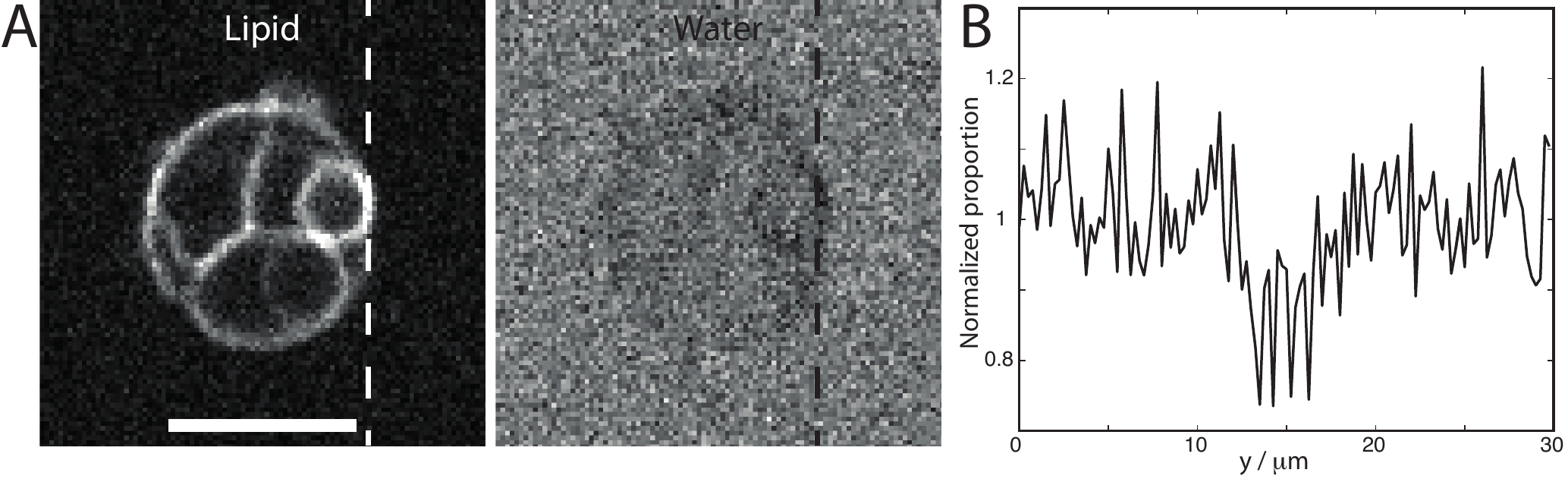}}
	\caption{
		High-sensitivity detection of lipid membranes in vesicles. 
		(A) Proportions of lipids (left panel) and water (right panel). The water channel has been scaled to enhance the appearance of the signal dip. Scale bar: 10~$\mu$m. Total acquisition time: 30~s. Effective pixel dwell time: 2.4~ms.
		(B) Normalized (by the image mean) line profiles of the water proportion. The signal dip is due to a depletion of water from the focal volume, which allows estimation of the number of bilayers in the focal volume (see text for further details).
	}
	\label{fig5}
\end{figure}

Finally, we use the compressive spectrometer to demonstrate sensitive imaging of biological specimens with sizes below the diffraction limit. The sample chosen are vesicles made of lipids as they are model systems in cell membrane biology~\cite{Ando2015,Donaldson2018}, and also because the eigenspectrum is not expected to considerably change for different environments. Fig.\ref{fig5}A shows the estimated proportions of the lipid and water response. In the lipid channel, there are clearly various bilayers wrapped around each other, a topology common to thin multilamellar vesicles~\cite{Moreaux2000}. In the water channel, there is a slight decrease in the overall water response due to the finite thickness of the lipid bilayers. The drop in intensity in the water channel allow us to estimate the water depletion thickness, as the Rayleigh length is much smaller than the vesicle curvature. From the intensity drop of the water channel (Fig. \ref{fig5}B), we estimate a maximum multilayer thickness of 54~nm (the strongest water proportion intensity dip), corresponding to 11 bilayers at that location.

\section*{Discussion}
We have presented a new spectrometer layout designed for single-pixel spectroscopy. With this layout, we showcase two types of compressive Raman modality, either without a priori knowledge or with a priori knowledge of the eigenspectra of the hyperspectrum. The main purpose of the high-throughput spectrometer is to enable sensitive measurements of specimens that are smaller than the microscope resolution; Because the solvent background generates a strong background, we reduce its intensity by tight focusing, and also increase the dynamic range of the detection system with low-noise and high-sensitivity detection.

Remarkably, we could detect with molecular selectivity few bilayers with the current design. Previously, vesicle systems were mostly imaged by coherent Raman scattering techniques~\cite{Mueller2002,Potma2003,Li2005,Cheng2003,Chen2010,Gasecka2017,Shen2017}, which are considerably more expensive and complex than the layout presented here. We estimated that the strongest region in Fig.~\ref{fig5}A contains about 11 bilayers. Given the linearity of the spontaneous Raman process on the local concentration, we expect single-bilayer~\cite{Ando2015,Rieger2017,Donaldson2018} imaging with slight improvements in the current design, for instance, by addition of a second detector~\cite{Rehrauer2018} and with larger clear-aperture gratings to reduce losses due to clipping.

There have been various demonstrations to accelerate Raman imaging for biological applications. These include line scan methods~\cite{Okada2012}, spatial focii multiplexing~\cite{Kong2016}, and widefield microscopy~\cite{Muller2016,Wadduwage2017}. High-sensitivity in slit-based spectrometers~\cite{Okada2012,Kong2016} is challenging as non-confocal geometries are hard to reach diffraction-limited performance~\cite{Schlucker2003}. In that sense, light-sheet excitation-widefield imaging~\cite{Muller2016} could probably provide highest sensitivity, however the whole image plane has to be optically clear and static during acquisition because of the spectral/temporal sampling. The abovementioned methods all use sensitive cameras, contraty to single-pixel approaches, thus avoiding slow and noisy measurements at high speeds (mostly read-out noise)~\cite{Scotte2018}. In addition, the combination of single-pixel-spectroscopy with tight-focusing (or confocal) allows for low-power CW lasers to be used. Nevertheless, the main disadvantage of compressive Raman with a priore information is its need for at least one long scanning to obtain the eigenspectra "dictionary", although a "blind" sampling approach to learn the dictionary has been proposed~\cite{Wilcox2013}. Note that even in biological applications one expects only a few eigenspectra~\cite{Klein2012}, which supports that compressive Raman approaches may have the fastest imaging speeds.

Previous compressive spectrometers used less sensitive detection. Quantum efficiency and dark counts of the detector are two key parameters for high-sensitivity detection as they determine the overall throughput and dynamic range for proportion estimation. The overall throughput scaling of the grating-based approach is $\phi_l^4~\phi_\text{\tiny DMD}~\phi_g^2~\phi_d$, where $\phi_i$ denotes the throughput of the lenses ($l$), DMD ($DMD$), gratings ($g$) and detector ($d$) (by detector throughput we mean quantum efficiency). Given that lenses and gratings may have negligible losses, the largest penalty is the DMD and detector quantum efficiency. In our design, we achieve a theoretical throughput of 34\% ($\phi_g=85\%$, $\phi_d=70\%$, $\phi_\text{\tiny DMD}=70\%$). A higher throughput value (50\%) is expected if operating in the near-IR as optical elements are commercially available. Note, however, while the sensitivity might be higher, the overall signal will be lower as the Raman cross-section scales as $\lambda^{-4}$. An alternative grating-based approach is to fold the design (back-reflect the light into the same grating)\cite{Davis2011}, however this would be even more lossy as the DMD has the lowest throughput in the design. Finally, the lens-based approach may reach overall throughput of 50\% after meticulous design of the optical elements.

\section*{Conclusion}
High-sensitivity Raman imaging means detecting faint signals over strong idle backgrounds. To address that, we have presented two inexpensive layouts of a high-throughput spectrometer for single-pixel compressive Raman microspectroscopy, and demonstrated its application with two different levels of a priori information. Such layout allowed for high-sensitivity imaging of few-bilayers vesicle in few seconds, therefore paving the way of compressive Raman for cell biology applications.

\section*{Acknowledgements}
We thank Tom Sperber for fruitful discussions. H.B.A. was supported by LabEX ENS-ICFP: ANR-10-LABX-0010/ANR-10-IDEX-0001-02 PSL*. This works was supported by European Research Council (ERC) (724473), Universitat Jaume I (PREDOC/2013/32), Generalitat Valenciana (PROMETEO/2016/079), Ministerio de Economia y Competitividad (MINECO, FIS2016-75618-R), Agence Nationale de la Recherche (ANR) France Bio Imaging (ANR-10-INSB-04-01) and France Life Imaging (ANR-11-INSB-0006) infrastructure networks and Plan cancer INSERM (PC201508). S.G. is a member of the Institut Universitaire de France.



%

\end{document}